\documentclass{article}
\usepackage[preprint]{spconf}
\usepackage{amsmath}
\usepackage{graphicx}
\usepackage{hyperref}
\usepackage{xspace}
\usepackage{amsmath}
\usepackage{amssymb}
\usepackage{graphicx} 
\usepackage{booktabs}
\usepackage{multirow}
\usepackage{comment}
\usepackage{arydshln}
\usepackage{xcolor}
\usepackage{multicol}
\usepackage{tabularray}
\usepackage{tabularx}
\usepackage[most]{tcolorbox}
\usepackage{enumitem}
\usepackage{fancyhdr}

\copyrightnotice{\copyright IEEE 2025}
\toappear{
    \begin{tabular}[t]{@{}l@{}}
        \copyright This work has been submitted to the IEEE for possible publication.\\Copyright may be transferred without notice, after which this version may no longer be accessible.
    \end{tabular}
}

\title{PhysHDR: When Lighting Meets Materials and Scene Geometry \\in HDR Reconstruction}

\name{Hrishav Bakul Barua$^{\ddagger}$$^{\dagger}$$^{\star}$, Kalin Stefanov$^{\dagger}$, Ganesh Krishnasamy$^{\ddagger}$, KokSheik Wong$^{\ddagger}$$^{\S}$, Abhinav Dhall$^{\dagger}$
\thanks{$^{\star}$This research is supported by the Global Excellence and Mobility Scholarship (GEMS), Monash University (Australia \& Malaysia).}
\thanks{$^{\S}$This research is supported, in part, by the E-Science fund under the project: \emph{Innovative High Dynamic Range Imaging - From Information Hiding to Its Applications} (Grant No. 01-02-10-SF0327).}
}

\address{
$^{\ddagger}$School of Information Technology, Monash University, Malaysia\\
$^{\dagger}$Faculty of Information Technology, Monash University, Australia\\
$^{\star}$Robotics and Autonomous Systems Lab, TCS Research, India\\
\{hrishav.barua, kalin.stefanov, ganesh.krishnasamy, wong.koksheik, abhinav.dhall\}@monash.edu}

\makeatletter
\DeclareRobustCommand\onedot{\futurelet\@let@token\@onedot}
\def\@onedot{\ifx\@let@token.\else.\null\fi\xspace}
\def\eg{\emph{e.g}\onedot} 
\def\ie{\emph{i.e}\onedot}

\makeatother

\let\oldthebibliography\thebibliography
\renewcommand\thebibliography[1]{%
  \oldthebibliography{#1}%
  \setlength{\itemsep}{0pt}   
  \setlength{\parskip}{0pt}  
}

\begin{document}

\maketitle

%
% Abstract.
\begin{abstract}
Low Dynamic Range (LDR) to High Dynamic Range (HDR) image translation is a fundamental task in many computational vision problems.
Numerous data-driven methods have been proposed to address this problem; however, they lack explicit modeling of illumination, lighting, and scene geometry in images.
This limits the quality of the reconstructed HDR images.
Since lighting and shadows interact differently with different materials, (\eg, specular surfaces such as glass and metal, and lambertian or diffuse surfaces such as wood and stone),
%Therefore, 
modeling material-specific properties (\eg, specular and diffuse reflectance) has the potential to improve the quality of HDR image reconstruction.
This paper presents PhysHDR, a simple yet powerful latent diffusion-based generative model for HDR image reconstruction.
The denoising process is conditioned on lighting and depth information and guided by a novel loss to incorporate material properties of surfaces in the scene. 
The experimental results establish the efficacy of PhysHDR in comparison to a number of recent state-of-the-art methods.
\end{abstract}

%
% Keywords.

\section{Introduction}
\label{sec:intro}
Reconstructing High Dynamic Range (HDR) images from Low Dynamic Range (LDR) counterparts has gained significant attention in the vision community~\cite{wang2021deep}.
Applications involving medical and computational imaging~\cite{lu2024pano}, robotic vision and self-driving cars~\cite{wu2020hdr}, augmented/virtual reality~\cite{singh24_hdrsplat}, media \& entertainment~\cite{he2022sdrtv}, require high-fidelity images of real-world scenes, which LDR images generally lack.
A large body of data-driven methods attempts to solve major issues in HDR imaging, pertaining to artifacts, ghosting, and blurring effects.
These methods primary approximate the reverse of the image formation pipeline in standard cameras, where a camera captures HDR scenes with high intensity values and clips them to a low dynamic range~\cite{liu2020single}.

\begin{figure}[t]
\centering
\includegraphics[width=\linewidth]{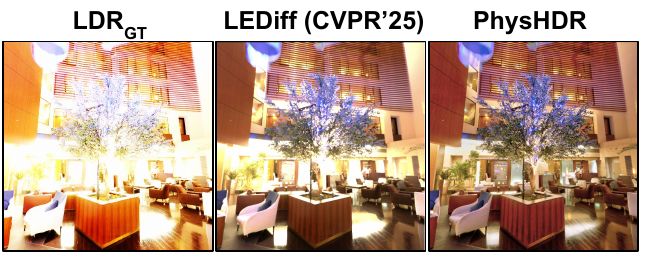}
\caption{PhysHDR (right) can recover the light/shadow details and light-object interactions in the rough and metallic surfaces better than the state-of-the-art~\cite{wang2025lediff} (middle) given an extremely over-exposed LDR image (left) as input.}
\label{fig:teaser}
\end{figure}

The current state-of-the-art focuses on the retrieval of information in low light areas~\cite{yang2023lightingnet} or extreme lighting conditions~\cite{nguyen2023psenet}.
Most methods utilize Convolutional Neural Networks (CNN)~\cite{shin2018cnn}, Transformers~\cite{liu2022ghost} and Generative Adversarial Networks (GAN)~\cite{nam2024deep,wang2023glowgan}.
Some methods use single-exposed LDR~\cite{barua2024histohdr,liu2020single,zou2023rawhdr}, while others use multi-exposed LDR images as input~\cite{barua2023arthdr,SelfHDR}.
More recent methods are based on diffusion models~\cite{wang2025lediff,yan2023toward,bemana2025bracket}, with some based on stable or latent diffusion and others on conditional diffusion~\cite{rombach2022high}.
These methods typically include a variational autoencoder (VAE)~\cite{kingma2019introduction} to encode and decode the input and output, a U-Net for the denoising process, and a condition encoder such as Contrastive Language-Image Pre-Training (CLIP)~\cite{radford2021learning} to embed the condition features. The denoising process takes place in the latent space instead of the pixel space. The main advantage of using diffusion over other generative-based or CNN-based methods is their ability to reconstruct high-resolution HDR images while mitigating artifacts and ghosting effects.

Although these methods reconstruct excellent HDR images, they lack explicit modeling of scene geometry (\ie, depth information), illumination conditions, and scene material properties. HDR reconstruction is an ill-posed problem because multiple real-world lighting conditions can result in the same LDR pixel values (especially in under- and over-exposed regions).
Therefore, it is technically challenging to disambiguate the pixel intensity values corresponding to the lighting and shadow areas in the reconstructed HDR.
To this end, we propose PhysHDR, a latent diffusion approach which harnesses the power of both depth and illumination information to reconstruct HDR images in a more physics-informed manner.
Inspired by the fact that different surfaces respond differently to light depending on their material properties (\eg, lambertian and specular), PhysHDR introduces a novel loss for material properties (\ie, albedo, roughness, and metallic) to guide the model in disambiguating the pixel intensities in the HDR image.
Lambertian (diffuse) materials such as wood and stone reflect light in all directions uniformly, in contrast, specular materials such as glass and metal reflect light only in one direction.
Depth combined with illumination information provides explicit 3D scene geometry, which helps to disentangle shading, lighting, and reflectance components.
Fig.~\ref{fig:teaser} illustrates the quality of HDR images reconstructed with PhysHDR compared to the recent state-of-the-art~\cite{wang2025lediff}.
Our \textbf{key-contributions} are as follows:
%\begin{itemize}[noitemsep,topsep=0pt]
\textbf{(a)} proposing a novel latent diffusion method conditioned on depth and illumination information to model the shading, lighting, and reflectance properties of materials in an unambiguous manner; 
\textbf{(b)} %\item{We 
proposing a new loss function based on material properties to further strengthen the efficacy in reconstructing light and shadow interactions with lambertian and specular surfaces, and;
%\item{We 
\textbf{(c)} presenting an extensive analysis of the method to highlight the contribution of each of the components.
%\end{itemize}

\section{Method}
\label{sec:method}
\begin{figure}[t]
\centering
\includegraphics[width=\linewidth]{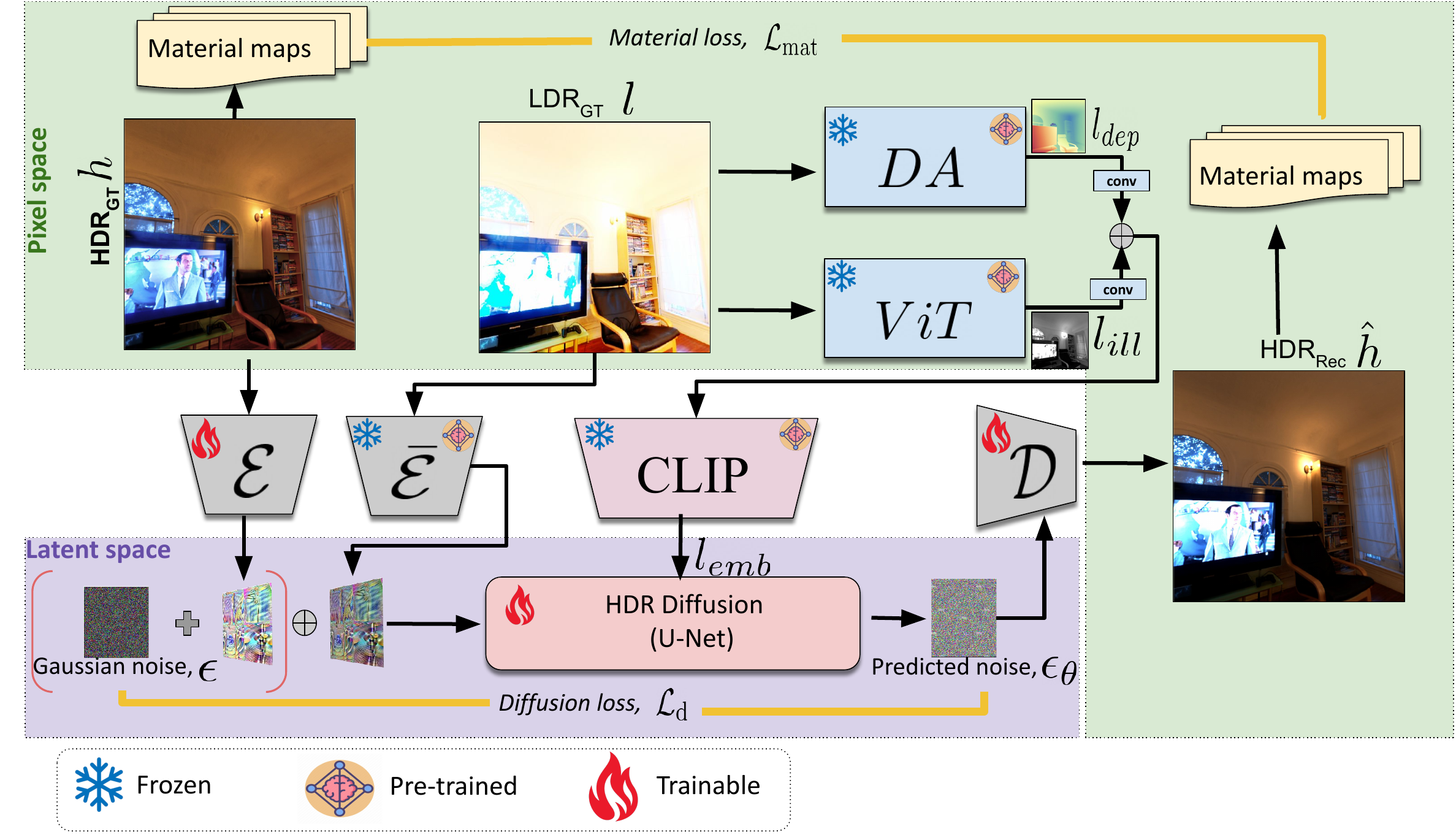}
\caption{Architecture of the proposed PhysHDR method.}
\label{fig:pipeline}
\end{figure}

The goal is to reconstruct an HDR image $\hat{h} \in \mathbb{R}^{H \times W \times 3}$ with $\gg 2^8$ radiance values (having luminance and color information for each pixel) given a single LDR image $l \in \mathbb{R}^{H \times W \times 3}$ with $2^8$ intensity values.
The proposed latent diffusion model PhysHDR, similar to~\cite{kocsis2024intrinsic,liu2023zero}, is conditioned on the input LDR image and its properties, \ie, illumination and scene geometry (depth maps containing surface normal information~\cite{ha2021normalfusion}) beneficial for learning light-object interactions.
The reconstructed HDR $\hat{h}$ and ground truth HDR $h$ are decomposed into three material maps (albedo, roughness, and metallic) using~\cite{kocsis2024intrinsic}, and employed in a novel loss function to further guide the diffusion process in reconstructing physics-informed light-object interactions.
To the best of our knowledge, PhysHDR (see Fig.~\ref{fig:pipeline}) is the first method to use material-based properties along with scene geometry (depth) and illumination information for HDR reconstruction.

\noindent\textbf{Architecture:}
PhysHDR uses stable diffusion U-Net~\cite{rombach2022high} as its base architecture.
% We fine-tune (some parts of) the pre-trained model with our data and additional conditioning information for our purpose.
Diffusion models have shown excellent performance as effective priors trained using huge amounts of real data~\cite{ho2020denoising,rombach2022high}.They transform Gaussian noise from the training distribution to data samples (\ie, $\hat{h} \sim q(h \mid l)$) using an iterative denoising process.
Similar to \cite{rombach2022high} we adopt the latent diffusion process with trainable encoder $\mathcal{E}(\cdot)$ (encodes the HDR image $h$ in latent representation), and decoder $\mathcal{D}(\cdot)$ (decodes latent representation to HDR image $\hat{h}$).
We use the LDR image $l$ as a condition in three different ways.
First, a pre-trained encoder $\mathcal{\bar{E}}(\cdot)$ (same architecture as $\mathcal{E}(\cdot)$) extracts features for $l$, which are concatenated with noise-induced features from $h$, making the input channel size six.
Second, illumination features $l_{ill}$ are extracted through the pre-trained $ViT$ encoder from \cite{Cui_2022_BMVC} and depth information $l_{dep}$ is extracted with the pre-trained model Depth Anything ($DA$)~\cite{depthanything}.
Finally, the outputs of $1\times1$ convolutional layers are concatenated (\ie, $l_{ill}$ $\oplus$ $l_{dep}$) and used in a pre-trained CLIP~\cite{radford2021learning} encoder to extract image embedding $l_{emb}$.
This embedding is used as a condition in the diffusion model's cross-attention U-Net during denoising.
The illumination and depth-aware features provide the diffusion process with scene geometry and disentangled lighting and shadow effects with various objects and surfaces.

\noindent\textbf{Diffusion:}
During the forward process, in each timestep $t \sim [1, 1000]$ a Gaussian noise $\epsilon$ $\sim$ $\mathcal{N}(\mu, \sigma^2)$ is added to $\mathcal{E}(h)$.
In the reverse process, $\epsilon_\theta(\cdot)$, an U-Net~\cite{rombach2022high}, predicts the added noise, where $\theta$ denotes the model parameters.
The denoising is conditioned on the features extracted from $l$ and the noisy $h$.
The training objective is:
\begin{align}
\label{eq:loss_1}
\small
\begin{split}
\mathcal{L}_{\text{d}} = 
\mathbb{E}_{h,\,\epsilon \sim \mathcal{N}(0,I),\,t}
\left[ \, \lVert \epsilon - \epsilon_\theta(\mathcal{E}(h) + \epsilon, t, \mathcal{\bar{E}}(l), l_{emb}) \rVert_2^2 \, \right]
\end{split}
\end{align}

During inference, we use the regular diffusion process~\cite{rombach2022high} to sample HDR features $\hat{h}$ conditioned on the LDR image $l$. The advantage of using a pre-trained diffusion prior lies in the fact that it has been trained using a huge volume of real-world high resolution data~\cite{rombach2022high,kocsis2024intrinsic}, \ie, the LDR encoder $\mathcal{\bar{E}}(\cdot)$ is frozen.
However, the HDR encoder $\mathcal{E}(\cdot)$ and decoder $\mathcal{D}(\cdot)$ are trained with our data.
Therefore, the process finetunes only selected modules as illustrated in Fig.~\ref{fig:pipeline}. 

\begin{table*}[t]
\setlength{\tabcolsep}{1pt}
\small
\centering
\caption{Intra-dataset comparison with the state-of-the-art. The best results are in \textcolor{red}{\textbf{bold}} and the second best are \textcolor{blue}{\underline{underlined}}.}
\label{Tab:quant}
\begin{tabular}{l|cccc|cccc}
\toprule
\multicolumn{1}{l|}{\multirow{2}{*}{\textbf{Method}}} & \multicolumn{4}{c|}{\textbf{City Scene}{~\cite{zhang2017learning}}} & \multicolumn{4}{c}{\textbf{HDR-Synth \& HDR-Real}{~\cite{liu2020single}}} \\
 & \textbf{PSNR}$\uparrow$ & \textbf{SSIM}$\uparrow$ & \textbf{LPIPS}$\downarrow$ & \textbf{VDP-3}$\uparrow$ &
\textbf{PSNR}$\uparrow$ & \textbf{SSIM}$\uparrow$ & \textbf{LPIPS}$\downarrow$ & \textbf{VDP-3}$\uparrow$ \\
\midrule[0.25mm]
Ghost-free{~\cite{liu2022ghost}} & \textcolor{red}{\textbf{40.11}}  & \textcolor{blue}{\underline{0.955}}   & 0.143  & 7.47  &  \textcolor{blue}{\underline{38.12}} &  \textcolor{blue}{\underline{0.943}} & 0.157 & 7.44 \\
GlowGAN{~\cite{wang2023glowgan}} & 34.01 & 0.902  & 0.167  & 7.41 & 33.10 & 0.901 & 0.171 & 7.32 \\
HistoHDR-Net{~\cite{barua2024histohdr}} & 35.14 & 0.940 & 0.311  & 7.50  & 33.48 & 0.910  & 0.342 & 7.34 \\
LEDiff{~\cite{wang2025lediff}} & 35.91 & 0.922  &  \textcolor{blue}{\underline{0.121}}  &  \textcolor{blue}{\underline{7.52}}  &  35.11 & 0.905 &  \textcolor{blue}{\underline{0.138}} &  \textcolor{blue}{\underline{7.48}} \\
\midrule
PhysHDR (Ours) &  \textcolor{blue}{\underline{39.01}} & \textcolor{red}{\textbf{0.971}}  & \textcolor{red}{\textbf{0.081}}  & \textcolor{red}{\textbf{7.85}}  & \textcolor{red}{\textbf{38.14}} & \textcolor{red}{\textbf{0.967}} & \textcolor{red}{\textbf{0.102}} & \textcolor{red}{\textbf{7.66}} \\
\bottomrule
\end{tabular}
\end{table*}

\noindent\textbf{Material loss:}
Apart from the diffusion loss in Eq.~\ref{eq:loss_1}, we also propose a novel objective function based on material properties of the objects and surfaces in the HDR images.
We first extract the material maps (\ie, albedo, roughness, and metallic) from the ground truth HDR $h$ and the reconstructed HDR $\hat{h}$ image using a state-of-the-art method which exhibits high accuracy in the task~\cite{kocsis2024intrinsic}.
We get $h_{al}$, $h_{ro}$, and $h_{met}$ from  $h$ and $\hat{h}_{al}$, $\hat{h}_{ro}$, and $\hat{h}_{met}$ from $\hat{h}$.
Albedo maps $\langle h_{al}$,$\hat{h}_{al} \rangle$ consists of the base color information of the objects in the scene, while roughness $\langle h_{ro}$,$\hat{h}_{ro} \rangle$ and metallic $\langle h_{met}$,$\hat{h}_{met}\rangle$ maps represent the degree of roughness and smoothness in any objects or surfaces.
While the diffusion loss ensures visual realism and produces attractive images, the material loss preserves physics-based properties of light and materials in the reconstructed HDR.
This loss is computed on the tone-mapped versions of the material maps.
This tone-mapping is performed using the $\mu$-law~\cite{jinno2011mu} and is done to avoid the high-intensity pixels of HDR images that can distort the loss calculation.
We define the loss between the material maps as: \begin{align}
\label{eq:mat_loss}
\small
\begin{split}
\mathcal{L}_{\text{mat}} = \frac{1}{N}
\sum_{n=1}^{N} \Bigl(\,\bigl\| h^{n}_{\text{al}} - \hat{h}^{n}_{\text{al}} \bigr\|_{1}
  + \,\bigl\| h^{n}_{\text{ro}} - \hat{h}^{n}_{\text{ro}} \bigr\|_{1}
  + \,\bigl\| h^{n}_{\text{met}} - \hat{h}^{n}_{\text{met}} \bigr\|_{1} \Bigr),
\end{split}
\end{align}
where N is the number of maps in each batch. 
The total objective of the model is: \begin{equation}
\label{eq:total_loss}
\mathcal{L_{\text{full}}} \;=\; \mathcal{L}_{d} \;+\; \lambda_{\text{mat}}\,\mathcal{L}_{\text{mat}},
\end{equation}
where $\lambda_{\text{mat}}$ is empirically set to 0.2 after experimenting with values from 0.1 to 0.5.

\section{Experiments and Results}
\label{sec:exp}
\noindent\textbf{Implementation:}
The finetuning of the pre-trained stable diffusion model~\cite{rombach2022high} (implemented in PyTorch) was done for 200 epochs with a batch size of 10, using AdamW optimizer~\cite{Loshchilov2017DecoupledWD} with a learning rate of $1e{-5}$.

\noindent\textbf{Datasets:}
To evaluate the performance of PhysHDR we used two datasets, including both real and synthetic images: City Scene dataset~\cite{zhang2017learning} (20K LDR/HDR image pairs) and HDR-Synth \& HDR-Real dataset~\cite{liu2020single} (9785 LDR/HDR real image pairs and around 500 synthetic pairs).
All images were resized to a resolution of $512\times512$.
We compared the performance of different methods in two experiments: intra- and cross-dataset evaluation.
For the intra-dataset experiment, we created 80\% train and 20\% test splits.
For the cross-dataset experiment, we considered an additional dataset, DrTMO~\cite{endo2017deep} (1043 LDR/HDR pairs).
For methods designed to use single-exposure LDR inputs, we provided one LDR image from the datasets that contain multiple exposures.
For methods that require multi-exposure LDR inputs (such as Ghost-free~\cite{liu2022ghost}), we synthetically generated (using the OpenCV function \texttt{convertScaleAbs}) the additional exposures for the datasets that contain only single-exposed LDR images.
In all ablation studies, we used the same test set sampled from the City Scene dataset~\cite{zhang2017learning}.

\begin{figure*}[t]
\centering
\includegraphics[width=\linewidth]{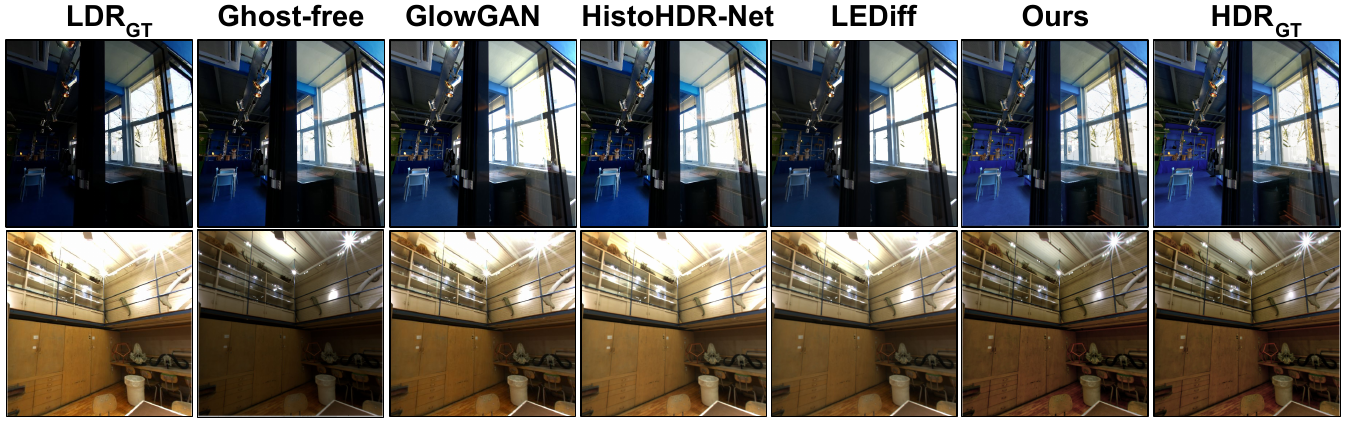}
\caption{HDR images reconstructed by the proposed PhysHDR and state-of-the-art methods. For our method, we specifically observe that the areas where light interacts with different surfaces are reconstructed realistically based on material properties.}
\label{fig:results}
\end{figure*}

\noindent\textbf{Metrics:}
We used four different metrics, \ie, Peak Signal-to-Noise Ratio in dB (PSNR), Structural Similarity Index Measure (SSIM)~\cite{wang2004image}, Learned Perceptual Image Patch Similarity (LPIPS), and High Dynamic Range Visual Differences Predictor (HDR-VDP-3)~\cite{cao2024perceptual}.
These metrics cover a wide range of evaluation parameters such as pixel-level similarity, structural similarity, and semantic and contextual similarity, and human-level perceptual judgement.
We calculated HDR-VDP-3, SSIM, and LPIPS scores using ground truth and reconstructed HDR images in the linear domain.
PSNR scores are obtained on $\mu$-law tone-mapped ground truth and reconstructed HDR images.

\noindent\textbf{Methods:}
We selected four different state-of-the-art methods: Ghost-free~\cite{liu2022ghost}, GlowGAN~\cite{wang2023glowgan}, HistoHDR-Net~\cite{barua2024histohdr}, and LEDiff~\cite{wang2025lediff} in our evaluations.
The selected methods include a CNN-based approach and generative approaches using GAN, Transformer, and Diffusion.

\noindent\textbf{Quantitative Results:}
We present the intra-dataset results in Table~\ref{Tab:quant}.
The proposed method outperforms the selected state-of-the-art methods in terms of PSNR, SSIM, LPIPS, and HDR-VDP-3 for HDR-Synth \& HDR-Real~\cite{liu2020single} and SSIM, LPIPS, and HDR-VDP-3 for City Scene~\cite{zhang2017learning}.
The Ghost-free~\cite{liu2022ghost} method outperforms our model in terms of PSNR on City Scene~\cite{zhang2017learning}.
LEDiff~\cite{wang2025lediff} performs second best for semantic similarity and human vision-based metrics while Ghost-free~\cite{liu2022ghost} performs second best for the pixel-based and structural similarity metrics.
The results of cross-dataset evaluation are presented in Table~\ref{tab:zero-shot}, and they show a similar trend as observed for  
%The results are similar to the 
intra-dataset evaluation, \ie, our method PhysHDR outperforms the state-of-the-art in all the metrices.

\begin{table}[t]
\setlength{\tabcolsep}{1pt}
\centering
\small
\caption{Cross-dataset evaluation of the proposed PhysHDR and state-of-the-art on unseen dataset (DrTMO~\cite{endo2017deep}). The best results are in \textcolor{red}{\textbf{bold}} and the second best \textcolor{blue}{\underline{underlined}}.}
\label{tab:zero-shot}
\begin{tabular}{l|cccc}
\toprule
\textbf{Method} & \textbf{PSNR}$\uparrow$ & \textbf{SSIM}$\uparrow$ & \textbf{LPIPS}$\downarrow$ & \textbf{VDP-3}$\uparrow$ \\
\midrule
Ghost-free{~\cite{liu2022ghost}} &  \textcolor{blue}{\underline{36.77}}  &  \textcolor{blue}{\underline{0.935}}   & 0.155   & 7.39\\
GlowGAN{~\cite{wang2023glowgan}}& 33.21 & 0.899   & 0.177  & 7.35\\
HistoHDR-Net{~\cite{barua2024histohdr}} & 33.43  & 0.908   & 0.351  & 7.41\\
LEDiff{~\cite{wang2025lediff}}& 34.12 &  0.909 &  \textcolor{blue}{\underline{0.142}}   &  \textcolor{blue}{\underline{7.52}}\\
\midrule
PhysHDR (Ours) &  \textcolor{red}{\textbf{37.89}}  & \textcolor{red}{\textbf{0.964}}  & \textcolor{red}{\textbf{0.109}}   & \textcolor{red}{\textbf{7.63}} \\
\bottomrule
\end{tabular}
\end{table}

\noindent\textbf{Qualitative Results:}
The visual quality of PhysHDR is illustrated in Fig.~\ref{fig:results}.
Our outputs closely resemble the ground truth HDR in terms of lighting and shadow quality as well as physically plausible light-object interactions on rough and metallic surfaces.
The base color (\ie, albedo) of the objects in the scene are also preserved with high fidelity.
We can also see the over-exposed and under-exposed areas with clarity.
We display the ground truth and generated HDR images using Reinhard's tone-mapping algorithm~\cite{10.1145/566654.566575}.

\noindent\textbf{Ablation Study:}
%\section{Ablation Study}
We performed ablation studies for architectural and loss components.
Table~\ref{tab:arch} summarizes the results, where the proposed components are added one by one, and to compare the improvement over the baseline model (\ie, U-Net-based denoising process with encoders $\mathcal{E}$ and $\mathcal{\bar{E}}$, and decoder $\mathcal{D}$).
The second row illustrates the contribution of CLIP embeddings extracted from the LDR $l$, leading to a significant improvement in all metrics.
The third and fourth rows illustrate the contribution of depth and illumination extracted from the LDR $l$, with small improvement for depth and a significant improvement for illumination.
The fifth row illustrates the contribution of combined depth and illumination resulting in improvement in all metrics.
The last row provides the results obtained when using all components of PhysHDR.
PhysHDR is trained with two objectives, $\mathcal{L}_{\text{d}}$ and $\mathcal{L}_{\text{mat}}$.
Table~\ref{tab:loss} provides an analysis of the contribution of the objectives.

\begin{table}[t]
\setlength{\tabcolsep}{1pt}
\centering
\small
\caption{Architecture ablation results for different components. The best results are in \textbf{bold}. Baseline: \cite{rombach2022high} + $\mathcal{E}$ + $\mathcal{\bar{E}}$, and; CLIP: CLIP embedding from $l$.}
\label{tab:arch}
\begin{tabular}{l|cccc}
\toprule
\textbf{Arch. components} & \textbf{PSNR}$\uparrow$ & \textbf{SSIM}$\uparrow$ & \textbf{LPIPS}$\downarrow$ & \textbf{HDR-VDP-3}$\uparrow$\\
\midrule
Baseline & 27.62 & 0.857  & 0.403 & 6.71 \\
+ CLIP & 29.23 & 0.897 & 0.307 & 6.84 \\
+ $l_{dep}$ & 29.11 & 0.891 & 0.288 & 6.91 \\
+ $l_{ill}$ & 33.12 & 0.912 & 0.276 & 7.21 \\
+ $l_{dep}$ $\oplus$ $l_{ill}$ & 34.12 & 0.934 & 0.126 & 7.43 \\
+ $l_{emb}$ & \textbf{39.01} & \textbf{0.971} & \textbf{0.081} & \textbf{7.85} \\
\bottomrule
\end{tabular}
\end{table}

\begin{table}[t]
\setlength{\tabcolsep}{1pt}
\centering
\small
\caption{Loss ablation results. The best results are in \textbf{bold}.}
\label{tab:loss}
\begin{tabular}{l|cccc}
\toprule
\textbf{Loss} & \textbf{PSNR}$\uparrow$ & \textbf{SSIM}$\uparrow$ & \textbf{LPIPS}$\downarrow$ & \textbf{HDR-VDP-3}$\uparrow$ \\
\midrule
$\mathcal{L}_{\text{d}}$ & 37.75 & 0.962 & 0.159 & 7.42 \\
$\mathcal{L}_{\text{d}}$ + $\mathcal{L}_{\text{mat}}$ & \textbf{39.01} & \textbf{0.971} & \textbf{0.081} & \textbf{7.85} \\
\bottomrule
\end{tabular}
\end{table}

\section{Conclusion}
Scene geometry (depth maps) and illumination information from the input image play a pivotal role in improving the performance of models for LDR to HDR reconstruction.
CLIP-based information from the input LDR further improves the efficacy.
The proposed material properties-based loss function ensures high-quality and perceptually realistic scene reconstruction, respecting the physics-based properties such as light-object and shadow-object interactions for different surfaces (\ie, metallic or rough).
Future work includes the study of diffusion models for HDR video reconstruction, as well as using material information directly as prior information in model training.
Another important consideration will be the use of normal maps explicitly in model training.

\vfill\pagebreak

\bibliographystyle{IEEEbib}
{\footnotesize
\bibliography{Bibliography}}

\begin{thebibliography}{10}

\bibitem{wang2021deep}
Lin Wang and Kuk-Jin Yoon,
\newblock ``{Deep Learning for HDR Imaging: State-of-the-Art and Future Trends},''
\newblock {\em IEEE transactions on pattern analysis and machine intelligence}, vol. 44, no. 12, pp. 8874--8895, 2021.

\bibitem{lu2024pano}
Zhan Lu, Qian Zheng, Boxin Shi, and Xudong Jiang,
\newblock ``Pano-nerf: Synthesizing high dynamic range novel views with geometry from sparse low dynamic range panoramic images,''
\newblock in {\em Proceedings of the AAAI Conference on Artificial Intelligence}, 2024, vol. 38-4, pp. 3927--3935.

\bibitem{wu2020hdr}
Xuesong Wu, Hong Zhang, Xiaoping Hu, Moein Shakeri, Chen Fan, and Juiwen Ting,
\newblock ``Hdr reconstruction based on the polarization camera,''
\newblock {\em IEEE Robotics and Automation Letters}, vol. 5, no. 4, pp. 5113--5119, 2020.

\bibitem{singh24_hdrsplat}
Shreyas Singh, Aryan Garg, and Kaushik Mitra,
\newblock ``Hdrsplat: Gaussian splatting for high dynamic range 3d scene reconstruction from raw images,''
\newblock {\em BMVC}, 2024.

\bibitem{he2022sdrtv}
Gang He, Kepeng Xu, Li~Xu, Chang Wu, Ming Sun, Xing Wen, and Yu-Wing Tai,
\newblock ``Sdrtv-to-hdrtv via hierarchical dynamic context feature mapping,''
\newblock in {\em Proceedings of the 30th ACM International Conference on Multimedia}, 2022, pp. 2890--2898.

\bibitem{liu2020single}
Yu-Lun Liu, Wei-Sheng Lai, Yu-Sheng Chen, Yi-Lung Kao, Ming-Hsuan Yang, Yung-Yu Chuang, and Jia-Bin Huang,
\newblock ``{Single-Image HDR Reconstruction by Learning to Reverse the Camera Pipeline},''
\newblock in {\em Proceedings of the IEEE/CVF Conference on Computer Vision and Pattern Recognition}, 2020, pp. 1651--1660.

\bibitem{wang2025lediff}
Chao Wang, Zhihao Xia, Thomas Leimkuhler, Karol Myszkowski, and Xuaner Zhang,
\newblock ``Lediff: Latent exposure diffusion for hdr generation,''
\newblock in {\em Proceedings of the Computer Vision and Pattern Recognition Conference}, 2025, pp. 453--464.

\bibitem{yang2023lightingnet}
Shaoliang Yang, Dongming Zhou, Jinde Cao, and Yanbu Guo,
\newblock ``Lightingnet: An integrated learning method for low-light image enhancement,''
\newblock {\em IEEE Transactions on Computational Imaging}, vol. 9, pp. 29--42, 2023.

\bibitem{nguyen2023psenet}
Hue Nguyen, Diep Tran, Khoi Nguyen, and Rang Nguyen,
\newblock ``Psenet: Progressive self-enhancement network for unsupervised extreme-light image enhancement,''
\newblock in {\em Proceedings of the IEEE/CVF Winter Conference on Applications of Computer Vision}, 2023, pp. 1756--1765.

\bibitem{shin2018cnn}
Seungjun Shin, Kyeongbo Kong, and Woo-Jin Song,
\newblock ``Cnn-based ldr-to-hdr conversion system,''
\newblock in {\em 2018 IEEE International Conference on Consumer Electronics (ICCE)}. IEEE, 2018, pp. 1--2.

\bibitem{liu2022ghost}
Zhen Liu, Yinglong Wang, Bing Zeng, and Shuaicheng Liu,
\newblock ``Ghost-free high dynamic range imaging with context-aware transformer,''
\newblock in {\em European Conference on Computer Vision}. Springer, 2022, pp. 344--360.

\bibitem{nam2024deep}
YoonChan Nam, JoonKyu Kim, Jae-hun Shim, and Suk-Ju Kang,
\newblock ``Deep conditional hdri: Inverse tone mapping via dual encoder-decoder conditioning method,''
\newblock {\em IEEE Transactions on Multimedia}, 2024.

\bibitem{wang2023glowgan}
Chao Wang, Ana Serrano, Xingang Pan, Bin Chen, Karol Myszkowski, Hans-Peter Seidel, Christian Theobalt, and Thomas Leimk{\"u}hler,
\newblock ``Glowgan: Unsupervised learning of hdr images from ldr images in the wild,''
\newblock in {\em Proceedings of the IEEE/CVF International Conference on Computer Vision}, 2023, pp. 10509--10519.

\bibitem{barua2024histohdr}
Hrishav~Bakul Barua, Ganesh Krishnasamy, KokSheik Wong, Abhinav Dhall, and Kalin Stefanov,
\newblock ``Histohdr-net: Histogram equalization for single ldr to hdr image translation,''
\newblock in {\em 2024 IEEE International Conference on Image Processing (ICIP)}. IEEE, 2024, pp. 2730--2736.

\bibitem{zou2023rawhdr}
Yunhao Zou, Chenggang Yan, and Ying Fu,
\newblock ``Rawhdr: High dynamic range image reconstruction from a single raw image,''
\newblock in {\em Proceedings of the IEEE/CVF International Conference on Computer Vision (ICCV)}, 2023, pp. 12334--12344.

\bibitem{barua2023arthdr}
Hrishav~Bakul Barua, Ganesh Krishnasamy, KokSheik Wong, Kalin Stefanov, and Abhinav Dhall,
\newblock ``{ArtHDR-Net: Perceptually Realistic and Accurate HDR Content Creation},''
\newblock in {\em 2023 Asia Pacific Signal and Information Processing Association Annual Summit and Conference (APSIPA ASC)}. IEEE, 2023, pp. 806--812.

\bibitem{SelfHDR}
Zhilu Zhang, Haoyu Wang, Shuai Liu, Xiaotao Wang, Lei Lei, and Wangmeng Zuo,
\newblock ``Self-supervised high dynamic range imaging with multi-exposure images in dynamic scenes,''
\newblock in {\em ICLR}, 2024.

\bibitem{yan2023toward}
Qingsen Yan, Tao Hu, Yuan Sun, Hao Tang, Yu~Zhu, Wei Dong, Luc Van~Gool, and Yanning Zhang,
\newblock ``Toward high-quality hdr deghosting with conditional diffusion models,''
\newblock {\em IEEE Transactions on Circuits and Systems for Video Technology}, vol. 34, no. 5, pp. 4011--4026, 2023.

\bibitem{bemana2025bracket}
Mojtaba Bemana, Thomas Leimk{\"u}hler, Karol Myszkowski, Hans-Peter Seidel, and Tobias Ritschel,
\newblock ``Bracket diffusion: Hdr image generation by consistent ldr denoising,''
\newblock in {\em Computer Graphics Forum}. Wiley Online Library, 2025, p. e70086.

\bibitem{rombach2022high}
Robin Rombach, Andreas Blattmann, Dominik Lorenz, Patrick Esser, and Bj{\"o}rn Ommer,
\newblock ``High-resolution image synthesis with latent diffusion models,''
\newblock in {\em Proceedings of the IEEE/CVF conference on computer vision and pattern recognition}, 2022, pp. 10684--10695.

\bibitem{kingma2019introduction}
Diederik~P Kingma, Max Welling, et~al.,
\newblock ``An introduction to variational autoencoders,''
\newblock {\em Foundations and Trends{\textregistered} in Machine Learning}, vol. 12, no. 4, pp. 307--392, 2019.

\bibitem{radford2021learning}
Alec Radford, Jong~Wook Kim, Chris Hallacy, Aditya Ramesh, Gabriel Goh, Sandhini Agarwal, Girish Sastry, Amanda Askell, Pamela Mishkin, Jack Clark, et~al.,
\newblock ``Learning transferable visual models from natural language supervision,''
\newblock in {\em International conference on machine learning}. PMLR, 2021, pp. 8748--8763.

\bibitem{kocsis2024intrinsic}
Peter Kocsis, Vincent Sitzmann, and Matthias Nie{\ss}ner,
\newblock ``Intrinsic image diffusion for indoor single-view material estimation,''
\newblock in {\em Proceedings of the IEEE/CVF Conference on Computer Vision and Pattern Recognition}, 2024, pp. 5198--5208.

\bibitem{liu2023zero}
Ruoshi Liu, Rundi Wu, Basile Van~Hoorick, Pavel Tokmakov, Sergey Zakharov, and Carl Vondrick,
\newblock ``Zero-1-to-3: Zero-shot one image to 3d object,''
\newblock in {\em Proceedings of the IEEE/CVF international conference on computer vision}, 2023, pp. 9298--9309.

\bibitem{ha2021normalfusion}
Hyunho Ha, Joo~Ho Lee, Andreas Meuleman, and Min~H Kim,
\newblock ``Normalfusion: Real-time acquisition of surface normals for high-resolution rgb-d scanning,''
\newblock in {\em Proceedings of the IEEE/CVF Conference on Computer Vision and Pattern Recognition}, 2021, pp. 15970--15979.

\bibitem{ho2020denoising}
Jonathan Ho, Ajay Jain, and Pieter Abbeel,
\newblock ``Denoising diffusion probabilistic models,''
\newblock {\em Advances in neural information processing systems}, vol. 33, pp. 6840--6851, 2020.

\bibitem{Cui_2022_BMVC}
Ziteng Cui, Kunchang Li, Lin Gu, Shenghan Su, Peng Gao, ZhengKai Jiang, Yu~Qiao, and Tatsuya Harada,
\newblock ``You only need 90k parameters to adapt light: a light weight transformer for image enhancement and exposure correction,''
\newblock in {\em 33rd British Machine Vision Conference 2022, {BMVC} 2022, London, UK, November 21-24, 2022}. 2022, {BMVA} Press.

\bibitem{depthanything}
Lihe Yang, Bingyi Kang, Zilong Huang, Xiaogang Xu, Jiashi Feng, and Hengshuang Zhao,
\newblock ``Depth anything: Unleashing the power of large-scale unlabeled data,''
\newblock in {\em CVPR}, 2024.

\bibitem{zhang2017learning}
Jinsong Zhang and Jean-Fran{\c{c}}ois Lalonde,
\newblock ``{Learning High Dynamic Range from Outdoor Panoramas},''
\newblock in {\em Proceedings of the IEEE International Conference on Computer Vision}, 2017, pp. 4519--4528.

\bibitem{jinno2011mu}
Takao Jinno, Hironori Kaida, Xinwei Xue, Nicola Adami, and Masahiro Okuda,
\newblock ``{$\mu$-Law Based HDR Coding and Its Error Analysis},''
\newblock {\em IEICE transactions on fundamentals of electronics, communications and computer sciences}, vol. 94, no. 3, pp. 972--978, 2011.

\bibitem{Loshchilov2017DecoupledWD}
Ilya Loshchilov and Frank Hutter,
\newblock ``Decoupled weight decay regularization,''
\newblock in {\em International Conference on Learning Representations}, 2017.

\bibitem{endo2017deep}
Yuki Endo, Yoshihiro Kanamori, and Jun Mitani,
\newblock ``{Deep reverse tone mapping},''
\newblock {\em ACM Trans. Graph.}, vol. 36, no. 6, pp. 177--1, 2017.

\bibitem{wang2004image}
Zhou Wang, Alan~C Bovik, Hamid~R Sheikh, and Eero~P Simoncelli,
\newblock ``Image quality assessment: from error visibility to structural similarity,''
\newblock {\em IEEE transactions on image processing}, vol. 13, no. 4, pp. 600--612, 2004.

\bibitem{cao2024perceptual}
Peibei Cao, Rafal~K Mantiuk, and Kede Ma,
\newblock ``Perceptual assessment and optimization of hdr image rendering,''
\newblock in {\em Proceedings of the IEEE/CVF Conference on Computer Vision and Pattern Recognition}, 2024, pp. 22433--22443.

\bibitem{10.1145/566654.566575}
Erik Reinhard, Michael Stark, Peter Shirley, and James Ferwerda,
\newblock ``Photographic tone reproduction for digital images,''
\newblock {\em ACM Trans. Graph.}, vol. 21, no. 3, pp. 267–276, jul 2002.

\end{thebibliography}

\end{document}